\documentstyle[12pt]{article}

\newcommand{\beq}{\begin{equation}}
\newcommand{\eeq}{\end{equation}}

\newcommand{\pa}{\partial}

\def\rlx{\relax\leavevmode}

\def\IZ{\rlx\hbox{\sf Z\kern-.4em Z}}
\def\IN{\rlx\hbox{\rm I\kern-.18em N}}
\def\IR{\rlx\hbox{\rm I\kern-.18em R}}
\def\IC{\rlx\hbox{\sf C\kern-.5em I }}

\newcommand{\dd}{\mbox{d}}

\newcommand{\su}{\mbox{s}}
\newcommand{\cc}{\mbox{c}}

\newcommand{\cx}{\IC [x_1,\ldots,x_N]}

\newcommand{\Dom}{<_{\mbox{\tiny D}}}
\newcommand{\Bru}{<_{\mbox{\tiny B}}}

\newcommand{\hD}{\widehat{D}}
\newcommand{\hatA}{\widehat{A}}
\newcommand{\hH}{\widehat{H}}

\newcommand{\hhB}{\widehat{h}^{(B)}}

\newcommand{\tilHb}{\widetilde{H}_B}
\newcommand{\calJ}{{\cal J}}
\newcommand{\hDelA}{\widehat{\Delta}^{(A)}}
\newcommand{\hDelB}{\widehat{\Delta}^{(B)}}

\newcommand{\Db}{D^{(B)}}
\newcommand{\hDa}{\widehat{D}^{(A)}}
\newcommand{\hDb}{\widehat{D}^{(B)}}

\newcommand{\Ab}{A}
\newcommand{\Abdag}{A^{\dag}}

\newcommand{\AnSB}{{\cal A}^{(B)}_{\mbox{s}}}
\newcommand{\AnCB}{{\cal A}^{(B)}_{\mbox{c}}}

\newcommand{\FsB}{{\cal F}^{(B)}_{\mbox{s}}}
\newcommand{\FcB}{{\cal F}^{(B)}_{\mbox{c}}}

\newcommand{\vacS}{|0\rangle_{\mbox{s}}}
\newcommand{\vacC}{|0\rangle_{\mbox{c}}}

\newcommand{\res}{\mbox{Res}}

\begin{document}

\title{An orthogonal basis for the $B_N$-type Calogero model}
\author{Saburo Kakei\thanks{E-mail: kakei@poisson.ms.u-tokyo.ac.jp}\\[5mm]
{\it Department of Mathematical Sciences, University of Tokyo,}\\
{\it  3-8-1 Komaba, Meguro-ku, Tokyo 153, Japan}}
\date{}

\maketitle

\begin{abstract}
We investigate algebraic structure for the $B_N$-type Calogero 
model by using the exchange-operator formalism. 
We show that the set of the Jack polynomials whose arguments 
are Dunkl-type operators provides an orthogonal basis. 
\end{abstract}
\bigskip
\bigskip

\section{Introduction}
Among quantum integrable models in one dimension, 
Calogero-Sutherland type models catch renewed interests
owing to the relation to fractional statistics.
An example of such models is the Calogero model with
harmonic potential \cite{Cal,Suth}:
\beq
H_A = \frac{1}{2}\sum_{j=1}^{N}
      \left( -\frac{\pa^2}{\pa x_j^2} + x_j^2 \right)
      + \sum_{j<k}\frac{\beta(\beta-1)}{(x_j-x_k)^2}.
\label{Ham:Cal}
\eeq
The subscript {\it ``A''} signifies that
this Hamiltonian is invariant under the action of the symmetric
group $S_N$, i.e. the $A_{N-1}$-type Weyl group.
There also exist Calogero-type models associated with other types
of the Weyl groups \cite{OP}.
The $B_N$-invariant counterpart of the Hamiltonian (\ref{Ham:Cal}) 
is the following \cite{Yam1,Yam2}:
\beq
H_B = \frac{1}{2}\sum_{j=1}^{N}
      \left\{ -\frac{\pa^2}{\pa x_j^2} + x_j^2 
             +\frac{\gamma(\gamma-1)}{x_j^2} \right\}
      + \sum_{j<k}\left\{
        \frac{\beta(\beta-1)}{(x_j-x_k)^2} 
        + \frac{\beta(\beta-1)}{(x_j+x_k)^2}\right\}.
\label{Ham:CalB}
\eeq
We remark that the model associated with the $C_N$-type Weyl group 
is equivalent to the $B_N$-case, and $D_N$-type model is obtained 
by setting $\gamma=0$.
The ground state wavefunction for this model is \cite{Yam1,Yam2}
\beq
\psi^{(B)}_0(x_1,\ldots,x_N) 
 = \prod_{j<k}|x_j^2 - x_k^2|^{\beta}
   \prod_{j=1}^N |x_j|^{\gamma}
   \prod_{j=1}^N \exp(-x_j^2/2).
\label{eq:gsCalB}
\eeq

Wavefunctions of the excited states are written as products
of $\psi^{(B)}_0$ and some symmetric polynomials.
Baker and Forrester obtained an orthogonal basis of such
polynomials and named ``generalized Laguerre polynomials'' 
\cite{BF,BF2}.
It should be noted that the properties of such
polynomials have been studied also by van Diejen \cite{vD}.
In \cite{BF}, the proof of the orthogonality is based on 
the orthogonality of another set of polynomials 
which they call ``generalized Jacobi polynomials''. 
They obtained the orthogonality of the generalized
Laguerre polynomials via some limiting procedure.

Here we make a kind of gauge-transformations on the 
Hamiltonian:
\begin{eqnarray}
\tilHb & = & (\phi_0^{(B)})^{-1} \circ H_B \circ \phi_0^{(B)}
\nonumber\\
 & = & \frac{1}{2}\sum_{j=1}^{N}
   \left( -\frac{\pa^2}{\pa x_j^2} + x_j^2 
          - \frac{2\gamma}{x_j}\frac{\pa}{\pa x_j}\right)
  - \beta \sum_{k\neq j}\frac{1}{x_j^2-x_k^2}
    \left(x_j\frac{\pa}{\pa x_j} - x_k\frac{\pa}{\pa x_k}\right),
\label{Ham:CalB1}
\end{eqnarray}
where $\phi_0^{(B)}$ is defined by
\beq
\phi^{(B)}_0(x_1,\ldots,x_N) =
\prod_{j<k}|x_j^2 - x_k^2|^{\beta}
        \prod_{j=1}^N |x_j|^{\gamma}.
\label{eq:partGS}
\eeq
To construct eigenstates of $\tilHb$, exchange-operator formalism
is also available \cite{Yam2}. One can construct an analogue of 
creation operators $\Abdag_j$ 
(for definition, see (\ref{eq:CrAn}) below)
and show that the wavefunctions of the form,
$
f\left( (\Abdag_1)^2,\ldots,(\Abdag_N)^2 \right)\prod_{j=1}^N 
\exp(-x_j^2/2),
$
become eigenstates of $\tilHb$ if $f(x_1,\ldots,x_N)$ are
homogeneous polynomials.
However naive choice of the polynomial
does not create the orthogonal states.
In previous work \cite{Kakei}, we have shown that the set of the 
Jack polynomials whose arguments are Dunkl-type operators provides 
an orthogonal basis for the $A_{N-1}$-type Calogero model.
The aim of this paper is to investigate the $B_N$-case.
We shall show that the Jack polynomials appear also in the $B_N$-case.

\section{Dunkl operators and Jack polynomials}
In this section, we briefly review the definition of the 
symmetric and non-symmetric Jack polynomials.
In physical context, the Jack polynomials appear as
polynomial part of wavefunctions for the Sutherland 
($1/\sin^2$-interaction) model.

We first introduce the Cherednik operators \cite{Ch,BGHP}:
\beq\label{op:ChereA}
\hDa_j = z_j\frac{\pa}{\pa z_j}
       + \beta \sum_{k(<j)}\frac{z_k}{z_j - z_k} (1-s_{jk})
       + \beta \sum_{k(>j)}\frac{z_j}{z_j - z_k} (1-s_{jk})
       + \beta (j-1) 
\eeq
where $s_{jk}$ are elements of the symmetric group $S_N$ 
(the $A_{N-1}$-type Weyl group).
An element $s_{ij}$ acts on functions of $z_1$, $\ldots$, $z_N$ 
as an operator which permutes arguments $z_i$ and $z_j$.
Since the operators $\hDa_j$ commute each other, they are diagonalized
simultaneously by suitable choice of bases of $\cx$ 
\cite{BGHP,Opdam}.
Such basis is called {\it non-symmetric} Jack polynomials.
An non-symmetric Jack polynomial $\calJ^{\lambda}_{w}(x)$,
labeled with the partition $\lambda=(\lambda_1,\ldots,\lambda_N)$ 
and the element $w\in S_N$, 
is characterized by the following properties \cite{BGHP,Opdam}:
\begin{enumerate}
\item $\displaystyle \calJ^{\lambda}_{w}(x) = 
      x^{\lambda}_{w} 
      + \sum_{(\mu,w')<(\lambda,w)}
        C^{\lambda \mu}_{w w'} x^{\mu}_{w'}$ ,
\item $\calJ^{\lambda}_{w}(x)$ is joint eigenfunctions for 
      the operators $\hDa_j$,
\end{enumerate}
where we have used the notation 
$x^{\lambda}_{w} = 
x^{\lambda_1}_{w(1)} \cdots x^{\lambda_N}_{w(N)}$.
To define the ordering $(\mu,w')<(\lambda,w)$, we use the
dominance ordering $\Dom$ for partitions \cite{Mac},
and the Bruhat ordering $\Bru$ for the elements of $S_N$ \cite{Hu}.
Using these, we define the ordering as follows:
\beq
(\mu,w')<(\lambda,w) \quad \Longleftrightarrow \quad
\left\{\begin{array}{ll}
(\mbox{i}) & \mu \Dom \lambda,\\
(\mbox{ii}) & \mbox{if } \mu = \lambda \mbox{ then }w'\Bru w.
\end{array}\right.
\eeq
We denote the eigenvalues of $\hDa_j$ as $\epsilon_j(\lambda,w)$:
\beq
\hDa_j \calJ^{\lambda}_w(x) = 
\epsilon_j(\lambda,w) \calJ^{\lambda}_w(x).
\label{eq:nonsymJ}
\eeq
The eigenvalues $\epsilon_j(\lambda,w)$ are all obtained
by permutating the components of the multiplet 
$\left\{ \lambda_{N-j+1} + \beta(j-1) \right\}_{j=1,\ldots,N}$.

Using the operator $\hDa_j$, we introduce generating function of 
symmetric commuting operators \cite{BGHP}:
\beq
\hDelA_{\su}(u) = \prod_{j=1}^N (u + \hDa_j).
\eeq
Since $\hDelA_{\su}(u)$ is symmetric in $\hD_j$, 
symmetric eigenfunctions are obtained by symmetrizing 
$\calJ^{\lambda}_{w}(x)$, 
which are nothing but the Jack symmetric polynomials $J_{\lambda}(x)$.
Eigenvalues of $\hDelA_{\su}(u)$ are then given by
\beq
\hDelA_{\su}(u) J_{\lambda}(x) = 
\prod_{j=1}^N
\left\{ u + \lambda_{N-j+1} + \beta(j-1) \right\}
J_{\lambda}(x).
\eeq
We note that all the eigenvalues of $\hDelA_{\su}(u)$ are distinct
for generic value of $u$.

We then introduce the $B_N$-type Dunkl operators \cite{Yam2,Dun}:
\beq
\Db_j = \frac{\pa}{\pa x_j}
  + \beta \sum_{k(\neq j)}\left\{
     \frac{1-s_{jk}}{x_j - x_k}
    + \frac{1-t_j t_k s_{jk}}{x_j + x_k} \right\}
  + \gamma \frac{1-t_j}{x_j}
\label{eq:B-Dunkl}
\eeq
where $s_{jk}$ and $t_j$ are elements of the $B_N$-type Weyl group.
An element $s_{ij}$ acts as same as in the $A_{N-1}$-case and 
$t_j$ acts as sign-change, i.e., replaces the coordinate $x_j$ by $-x_j$.
Commutation relations of the $B_N$-type Dunkl operators are
\beq
\begin{array}{l}
\displaystyle
[ \Db_i, \Db_j] = 0,\\
\displaystyle
[ \Db_i, x_j ] = 
      \delta_{ij}\left(1+ \beta
         \sum_{k(\neq j)}(s_{jk}+t_j t_k s_{jk})
         + 2\gamma t_j \right)
      -(1-\delta_{ij})\beta (s_{ij}-t_i t_j s_{ij}),\\
s_{ij}\Db_j = \Db_i s_{ij}, \qquad
s_{ij} \Db_k = \Db_k s_{ij} \quad ( k\neq i,j ),\\
t_j\Db_j = -\Db_j t_j, \qquad
t_j \Db_k = \Db_k t_j \quad ( k\neq j ).
\label{eq:CRdxB}
\end{array}
\eeq
We denote the algebra generated by the elements $x_j$, $\Db_j$,
$s_{ij}$ and $t_j$ as $\AnSB$.
We introduce an $\AnSB$-module $\FsB$ 
(``Fock space'' for $\AnSB$)
generated by the vacuum vector $\vacS =1$:
\beq
\FsB = \IC [x_1^2,\ldots,x_N^2] \vacS.
\eeq
The elements $\Db_j$ of $\AnSB$ annihilate the vacuum vector, 
and $s_{ij}$, $t_j$ preserve $\vacS$:
\beq
D_j \vacS = 0,\qquad s_{ij} \vacS = \vacS,
\qquad t_j \vacS = \vacS .
\label{eq:Dvac}
\eeq

We then define Cherednik-type commuting operators 
associated with (\ref{eq:B-Dunkl}):
\begin{eqnarray}
\hDb_j & = & x_j \Db_j 
       + \beta \sum_{k(<j)}(s_{jk} + t_j t_k s_{jk})\nonumber\\
 & = & x_j\frac{\pa}{\pa x_j}
 + \beta \sum_{k(<j)}\left\{
 \frac{x_k}{x_j - x_k}(1-s_{jk})-\frac{x_k}{x_j + x_k}(1-t_j t_k s_{jk})
 \right\}\nonumber\\
 & & \quad + \; \beta \sum_{k(>j)}\left\{
 \frac{x_j}{x_j - x_k}(1-s_{jk})+\frac{x_j}{x_j + x_k}(1-t_j t_k s_{jk})
 \right\} + 2\beta (j-1) + \gamma (1-t_j) .
\label{op:ChereB}
\end{eqnarray}
We introduce the notation $\res^{(t)}(X)$ which means the
action of the operator $X$ is restricted to the functions with 
the symmetry $t_j f(x) = f(x)$.
Under this restriction, the action of the operator $\hDb_j$ is 
reduced to the following form:
\begin{eqnarray}
\res^{(t)}(\hDb_j) & = & x_j\frac{\pa}{\pa x_j}
  + 2\beta \sum_{k(<j)}\frac{x_k^2}{x_j^2 - x_k^2}(1-s_{jk})
\nonumber\\
& & \qquad + 2\beta \sum_{k(>j)}\frac{x_j^2}{x_j^2 - x_k^2} (1-s_{jk})
  + 2\beta (j-1).
\label{op:ResChereB}
\end{eqnarray}
Comparing (\ref{op:ResChereB}) with (\ref{op:ChereA}), we find that
$\res^{(t)}(\hDb_j)$ is equivalent to $2\hDa_j$ if we make a change 
of the variables $z_j=x_j^2/2$.
If we define the operator $\hDelB_{\su}(u)$ as
\beq
\hDelB_{\su}(u) = \prod_{j=1}^N (u + \hDb_j),
\eeq
we have the following equation by using the correspondence 
between $\res^{(t)}(\hDb_j)$ and $2\hDa_j$:
\begin{eqnarray}
\lefteqn{\hDelB_{\su}(u) 
J_{\lambda} \left(x_1^2/2,\ldots,x_N^2/2 \right) }\quad\nonumber\\
& = & \prod_{j=1}^N
\left\{ u + 2\lambda_{N-j+1} + 2\beta(j-1) \right\}
J_{\lambda}\left( x_1^2/2,\ldots,x_N^2/2 \right).
\label{eq:eigenSuthB}
\end{eqnarray}

\section{$B_N$-type Calogero model}
We now turn to the $B_N$-type Calogero model.
We introduce an analogue of creation and annihilation
operators \cite{Yam2}:
\beq
\Abdag_j = \frac{1}{\sqrt{2}} (-\Db_j + x_j), \qquad
\Ab_j = \frac{1}{\sqrt{2}} (\Db_j + x_j).
\label{eq:CrAn}
\eeq
By a direct calculation, we can show that
the operator $\Abdag_j$ is adjoint of 
$\Ab_j$ with respect to the scalar product
\beq
(f,g)_B = \int_{-\infty}^{\infty}
f(x_1,\ldots,x_N)g(x_1,\ldots,x_N)(\phi^{(B)}_0)^2
\prod_{j=1}^N \mbox{d} x_j .
\label{eq:SP2}
\eeq

We call an algebra generated by $\Ab_j$, $\Abdag_j$, $s_{ij}$
and $t_j$ as $\AnCB$. 
Since the commutation relations of these operators are the same as
those of $x_j$ and $\Db_j$,
we can define an isomorphism of $\AnSB$ to $\AnCB$ as follows:
\beq
\sigma (x_j) = \Abdag_j, \qquad \sigma (\Db_j) = \Ab_j .
\eeq
Fock space for $\AnCB$ is constructed in the same way as $\FsB$:
\beq
\FcB = \IC [(\Abdag_1)^2,\ldots,(\Abdag_N)^2] \vacC,
\eeq
with $\vacC = \prod_{j=1}^N \exp(-x_j^2/2)$.
The elements $\Ab_j$ of $\AnCB$ annihilate the vacuum vector, 
and $s_{ij}$, $t_j$ preserve $\vacC$:
\beq
\Ab_j \vacC = 0,\qquad s_{ij} \vacC = \vacC,
\qquad t_j \vacC = \vacC .
\label{eq:Avac}
\eeq
Comparing (\ref{eq:Avac}) with (\ref{eq:Dvac}), we know that 
the isomorphism $\sigma$ can be extended to the isomorphism 
of the Fock spaces:
\beq
\sigma(\vacS) = \vacC, \qquad
\sigma(a |v\rangle) = \sigma(a)\sigma(|v\rangle)
\label{eq:iso}
\eeq
for $a\in\AnSB$ and $|v\rangle\in\FsB$.

Applying this isomorphism to (\ref{eq:eigenSuthB}), we get the 
following equation:
\begin{eqnarray}
\lefteqn{\hDelB_{\cc}(u) J_{\lambda}
\left((\Abdag_1)^2/2,\ldots,(\Abdag_N)^2/2\right)
\vacC}\quad\nonumber\\
 & = & \prod_{j=1}^N
\left\{ u + 2\lambda_{N-j+1} + 2\beta(j-1) \right\}
J_{\lambda}
\left((\Abdag_1)^2/2,\ldots,(\Abdag_N)^2/2\right)
\vacC,
\label{eq:eigenCalB}
\end{eqnarray}
where we define $\hDelB_{\cc}(u)$ as
\beq
\hDelB_{\cc}(u) = \sigma\left( \hDelB_{\su}(u) \right)
 = \prod_{j=1}^N (u + \hhB_j)
\label{eq:CQb}
\eeq
with
\beq
\hhB_j = \sigma\left( \hDb_j \right)
 = \Abdag_j \Ab_j 
       + \beta \sum_{k(<j)}(s_{jk} + t_j t_k s_{jk}).
\eeq
We note that the operator $\hhB_j$ is self-adjoint with respect 
to (\ref{eq:SP2}).
The transformed Hamiltonian $\tilHb$ is related to 
(\ref{eq:CQb}) as follows;
if we denote the $(N-j)$-th coefficient of $\hDelB_{\cc}(u)$ as 
$I^{(B)}_{\cc,j}$, then $\tilHb$ is obtained from
$I^{(B)}_{\cc,1}$ after restricting to the $B_N$-invariant 
subspace:
\beq
\res\left(I^{(B)}_{\cc,1}\right)
= \res\left(\sum_{j=1}^N \hhB_j \right)
= \tilHb -\frac{N}{2} - \gamma N.
\eeq

From (\ref{eq:eigenCalB}) we find that all the eigenvalues of 
$\hDelB_{\cc}(u)$ are distinct.
On the other hand, 
the operator $\hDelB_{\cc}(u)$ is self-adjoint 
with respect to the scalar product (\ref{eq:SP2}).
From these facts, we conclude that the vectors,
\beq
|\lambda\rangle = J_{\lambda}
\left( (\Abdag_1)^2/2,\ldots,(\Abdag_N)^2/2 \right)\vacC,
\label{eq:OrthB}
\eeq
form an orthogonal basis with respect to 
the scalar product (\ref{eq:SP2}).
We note that polynomial parts of the basis (\ref{eq:OrthB})
are equivalent to the generalized Laguerre polynomials
introduced by Baker and Forrester up to constant.

Baker and Forrester also introduced ``non-symmetric generalized 
Laguerre polynomials'' \cite{BF2}.
In our formulation, such polynomials are related to joint 
eigenfunctions of the operators $\hhB_j$, which are of the form,
$
\calJ^{\lambda}_w
\left((\Abdag_1)^2/2,\ldots,(\Abdag_N)^2/2\right)\vacC .
$
The non-symmetric generalized Laguerre polynomials are 
polynomial parts of these eigenfunctions.

In conclusion, we have constructed operator expression of 
the orthogonal basis for the $B_N$-type Calogero model
by using the Jack polynomials whose arguments are the
Dunkl-type creation and annihilation operators.
We stress that our proof of orthogonality is algebraic
 and does not make use of the limiting procedure.

\section*{Appendix}
\renewcommand{\theequation}{\mbox{A}\arabic{equation}}
\setcounter{equation}{0}

In this appendix, we investigate the one-variable case 
in more detail to clarify the relationship to 
the Laguerre polynomials.

For $N=1$ case, Hamiltonian (\ref{Ham:CalB}) is reduced to 
\beq
\label{Ham:BOne}
\hH = \frac{1}{2}\left\{ -\frac{\dd^2}{\dd x^2} 
      + x^2 + \frac{\gamma(\gamma-1)}{x^2} \right\}.
\eeq
The ground state wavefunction is 
$\psi_0(x) = |x|^{\gamma}\exp(-x^2/2)$ whose eigenvalue is
$1/2 + \gamma$ (we omit the normalization constant).

On the other hand, the creation and annihilation operators 
(\ref{eq:CrAn}) are reduced to
\beq
\label{eq:CrAnOne}
A^{\dag} = \frac{1}{\sqrt{2}} \left\{
-\frac{\dd}{\dd x} + x - \frac{\gamma}{x}(1-\hat{t})\right\}, 
\qquad
A = \frac{1}{\sqrt{2}} \left\{
\frac{\dd}{\dd x} + x + \frac{\gamma}{x}(1-\hat{t})\right\},
\eeq
where $\hat{t}$ is the reflection operator $\hat{t}f(x)=f(-x)$.
The Hamiltonian (\ref{Ham:BOne}) and the 
operators (\ref{eq:CrAnOne}) are related as follows:
\[
\hH = |x|^{\gamma}\circ \frac{1}{2}
\res (A^{\dag}A +A A^{\dag})
\circ |x|^{-\gamma} ,
\]
where $\res\:X$ means that action of the operator $X$ is 
restricted to evne functions.

Wavefunctions for excited states can be constructed
by using gauge-transformed version of (\ref{eq:CrAnOne}), i.e.
\begin{eqnarray*}
\hatA^{\dag} &=& |x|^{\gamma} \circ A^{\dag} \circ |x|^{-\gamma}
= \frac{1}{\sqrt{2}}
\left(-\frac{\dd}{\dd x} + x + \frac{\gamma}{x}\,\hat{t}\right), \\
\hatA &=& |x|^{\gamma} \circ A \circ |x|^{-\gamma}
= \frac{1}{\sqrt{2}}
\left(\frac{\dd}{\dd x} + x - \frac{\gamma}{x}\,\hat{t}\right),
\end{eqnarray*}
which obey the commutation relations
\beq
\label{eq:CROne}
[ \hH, \hatA^{\dag} ] = \hatA^{\dag}, \qquad
[ \hH, \hatA ] = -\hatA.
\eeq
It should be noted that these creation and annihilation operators
have been introduced by Yang \cite{Yang}.

To preserve the symmetry $\psi(x)=\psi(-x)$, we apply $\hatA^{\dag}$
by even times to the ground state wavefunction $\psi_0(x)$:
\[
\psi_{2n}(x)= (\hatA^{\dag})^{2n} \psi_0(x).
\]
This formula is $N=1$ counterpart of (\ref{eq:OrthB}).
From (\ref{eq:CROne}), it follows that $\psi_{2n}(x)$ are also 
eigenfunctions of $\hH$:
\beq
\label{eq:Sch}
\hH \psi_{2n}(x) = (2n+1/2)\psi_{2n}(x).
\eeq
The wavefunctions $\psi_{2n}(x)$ are expressed as product of some 
polynomials $f_{2n}(x)$ and the ground state wavefunction $\psi_0(x)$.
Rewriting (\ref{eq:Sch}), one can obtain the following 
differential equation for $f_{2n}(x)$:
\[
\frac{\dd^2 f_{2n}}{\dd x^2} 
-\left( 2x-\frac{2\gamma}{x}\right)\frac{\dd f_{2n}}{\dd x} 
+ 4n f_{2n} =0.
\]
Making a change of the variable $y=x^2$, we obtain
\[
y\frac{\dd^2 f_{2n}}{\dd y^2} 
+\left(\frac{1}{2}+\gamma-y\right)\frac{\dd f_{2n}}{\dd y} 
+ n f_{2n} =0,
\]
which coincides with the differential equation for the 
Laguerre polynomials. 
Hence we conclude that $f_{2n}(x)$ can be written by using 
the Laguerre polynomials:
\[
f_{2n}(x) = n! (-2)^n L^{(\gamma-1/2)}_n(x^2).
\]

Since $\psi_{2n}(x)$ are even functions, the operators
$(\hatA^{\dag})^2$ and $\hatA^2$ act equivalently to 
\begin{eqnarray*}
B^{+} &=& \res ((\hatA^{\dag})^2) 
= \frac{1}{2}\left\{ \frac{\dd^2}{\dd x^2} -2x\frac{\dd}{\dd x}
  +x^2 -1 -\frac{\gamma(\gamma-1)}{x^2} \right\},\\ 
B^{-} &=& \res (\hatA^2) 
= \frac{1}{2}\left\{ \frac{\dd^2}{\dd x^2} +2x\frac{\dd}{\dd x}
  +x^2 +1 -\frac{\gamma(\gamma-1)}{x^2} \right\},
\end{eqnarray*}
respectively. 
We remark that the operators $B^+$ and $B^-$ have been introduced
by Perelomov \cite{Pere}.

The operators $B^+$ and $B^-$ give the recursion relations for
the wavefunctions:
\beq
\label{eq:rec}
B^+ \psi_{2n} = \psi_{2n+2}, \qquad
B^- \psi_{2n} = 4n\left(n -\frac{1}{2} +\gamma\right)\psi_{2n-2},
\eeq
where the constant factor of the second relation is determined
by comparing the coefficient of $x^{2n-2}\psi_0(x)$.
One can obtain recursion relations for the Laguerre polynomials
by rewriting (\ref{eq:rec}) \cite{Pere}:
\begin{eqnarray*}
&& \left\{ y\frac{\dd^2}{\dd y^2} 
  + \left(\frac{1}{2}+\gamma -2y\right)\frac{\dd}{\dd y} 
  + 2y -\frac{1}{2}-\gamma \right\} L^{(\gamma-1/2)}_n (y)
 = -(n+1) L^{(\gamma-1/2)}_{n+1} (y),\\
&& \left\{ y\frac{\dd^2}{\dd y^2} 
  + \left(\frac{1}{2}+\gamma\right)\frac{\dd}{\dd y} 
  \right\} L^{(\gamma-1/2)}_n (y)
 = -\left(n-\frac{1}{2}+\gamma\right) L^{(\gamma-1/2)}_{n-1} (y).
\end{eqnarray*}

\end{document}